\documentclass[12pt]{article}
\usepackage{amsmath}
\usepackage{amssymb}
\usepackage{epsfig}
\usepackage{euscript}
\usepackage{fancybox}
\usepackage{color}
\def\Color#1{\color[named]{#1}}

\def\mathswitchr#1{\relax\ifmmode{\mathrm{#1}}\else$\mathrm{#1}$\fi}

\newcommand{\umf}{{\Color{PineGreen}\mathfrak{u}}}
\newcommand {\pslash}{\hbox{$\not\hbox{\kern-2.3pt $p$}$}}

\usepackage{cite}

\usepackage{epic}
\def\alf1{ {\alpha\over\pi} }
\begin{document}
%\input{feynman} 
%=======================================================================
\begin{titlepage}
\begin{flushleft}
%{\bf MPI-PhT-2002-08}\\
{\bf BU-HEPP-10-07}\\
{\bf CERN-PH-TH/2011-029}\\
{\bf Jan., 2011}
\end{flushleft}
\vspace{0.05cm}
 
\begin{center}
%{\Large Chinese Magic in Loop Integrals$^{\dagger}$}
{\Large Magic Spinor Product Methods in Loop Integrals$^{\dagger}$}
\end{center}

\vspace{2mm}
\begin{center}
%%  {\bf   S. Jadach$^{a,b}$ and B.F.L. Ward$^{c,d}$}
{\bf   B.F.L. Ward}\\
\vspace{2mm}
%{\em $^a$CERN, Theory Division, CH-1211 Geneva 23, Switzerland,}\\
%{\em $^b$Institute of Nuclear Physics,
%        ul. Kawiory 26a, Krak\'ow, Poland,}
%{\em $^c$Werner-Heisenberg-Institut, Max-Planck-Institut fuer Physik,
%Muenchen, Germany,}\\
%{\em $^a$Werner-Heisenberg-Institut, Max-Planck-Institut fuer Physik,
%Muenchen, Germany,}\\
{\em Department of Physics,}\\
{\em Baylor University, Waco, Texas, USA}\\
{\em and}\\
{\em CERN, Geneva, Switzerland}\\
%{\em $^c$SLAC, Stanford University, Stanford, California 94309, USA,}\\
%{\em $^b$Department of Physics and Astronomy,\\
%  The University of Tennessee, Knoxville, Tennessee 37996-1200, USA}\\
%{\em $^c$SLAC, Stanford University, Stanford, California 94309, USA,}\\
\end{center}

\vspace{5mm}
\begin{center}
{\bf   Abstract}
\end{center}
We present an approach to higher point loop integrals using Chinese magic in the virtual loop integration variable. We show, using the five point function in the important $e^+e^-\rightarrow f\bar{f}+\gamma$ process for ISR as a pedagogical vehicle, 
that we get an expression for it directly
reduced to one scalar 5-point function and  4-, 3-, and 2- point integrals,
thereby avoiding the computation of the usual three tensor 
5-pt Passarino-Veltman 
reduction. We argue that this offers potential for greater numerical stability. 
\\
\vskip 20mm
\vspace{10mm}
\renewcommand{\baselinestretch}{0.1}
\footnoterule
\noindent
{\footnotesize
\begin{itemize}
\item[${\dagger}$]
Work partly supported by US DOE grant DE-FG02-09ER41600. 
% the Polish Government
%grants KBN 2P30225206 and 2P03B17210, the Maria Sk\l{}odowska-Curie
%Joint Fund II PAA/DOE-97-316, and
%by NATO Grant PST.CLG.980342.
%, and by
%Polish Government grant 5P03B09320.
\end{itemize}
}
%\vspace{0.5cm}
%\begin{flushleft}
%{\bf UTHEP-00-0101}\\
%{\bf Jan, 2000}\\
%\end{flushleft}

\end{titlepage}

%=======================================================================
\def\Kmax{K_{\rm max}}\def\ieps{{i\epsilon}}\def\rQCD{{\rm QCD}}
\renewcommand{\theequation}{\arabic{equation}}
\font\fortssbx=cmssbx10 scaled \magstep2
\renewcommand\thepage{}
%\vfill\eject
\parskip.1truein\parindent=20pt\pagenumbering{arabic}\par

%\section{\bf Introduction}\label{intro}\par
With the advent of the LHC, we enter the era of precision QCD, by which we mean
predictions for QCD processes at the total precision tag of $1\%$ or better.
This is analogous to the per mille level era of EW corrections at
LEP energies. Radiative effects at the level of ${\cal O}(\alpha_s^2)$ 
have to be controlled on the QCD side and those at the level of ${\cal O}(\alpha L \alpha_s),\;{\cal O}(\alpha^2L^2)$ on the QED$\otimes$QCD and QED sides
have to be controlled systematically, both from the physical precision standpoint and from the technical precision standpoint, in order to optimize
physics discovery at the LHC\footnote{Here, $L$ denotes the typical big log for the process under discussion.}. In Ref.~\cite{qced}, we have developed
a platform for the realization of such corrections ultimately on an event-by-event basis based on exact, amplitude-based resummation of QED and QCD together,
wherein residuals for hard photons and hard gluons are simultaneously
calculated order-by-order in perturbation theory in powers of
$\alpha$ and $\alpha_s$. These residuals, which are infrared finite and,
for hadron-hadron applications, collinearly finite require then exact
evaluation of higher point and (higher) loop
Feynman diagrams in an appropriate
reduction scheme for any attendant tensor properties as first 
developed systematically in Ref.~\cite{pass-velt}, for example.
Recently, alternative approaches have been developed 
in Refs.~\cite{z-dix-kos,opp} to deal with the growing complexity of the 
method in Ref.~\cite{pass-velt} as the number of legs beyond four
and/or loops beyond one
increases. 
Here, we focus on higher point one-loop functions\footnote{See Refs.~\cite{hypergm} 
for some recent progress on the higher loop functions
with an eye toward their use in the MC realization of the approach in Ref.~\cite{qced}}.\par
It has been demonstrated that n-point functions, for  $n=1,\cdots,4$, at one-loop,
reduced to scalar functions using the method of Ref.~\cite{pass-velt}, are tractable for fast MC event generator implementation for arbitrary
masses and kinematics for high energy scattering processes~\cite{bkj,bkh,alibaba,sjward,ceex,bdp,bdh,bvnb,bmr,bardin,hollik,bwl,fleis-jeg,bohm,melles,mcnlo,powheg}. It has also been
demonstrated~\cite{melrose,nto4-1,nto4-2,nto4-3} that, at one-loop, higher point scalar
functions can be
reduced to sums of four-point scalar functions.
In Refs.~\cite{nto4-1,oldenbgh,denner,dao,den-ditt} 
representations of the scalar four-point function that cover
arbitrary masses and the momenta relevant to most high energy
collider applications
have been given and these are suitable for fast MC implementation.
Thus, when one is discussing higher point functions at one-loop, we can consider,
at least for most collider physics applications, that the 1, 2, 3 and 4 point functions at one-loop are known in a practical way so that the main issue
can be considered to be 
the representation of the higher point functions in terms of these known functions. 
\par
When we consider any higher point function,
two of the most important aspects of any reduction procedure for
recasting it in terms of the ``known'', lower point functions are 
its numerical stability
and its usefulness for Monte Carlo event generator realization, as we 
have in mind for our residuals $\hat{\tilde{\bar\beta}}_{n,m}$ in Refs.~\cite{qced}
for example. Given the simplification that has been shown
for the ``Chinese magic'' 
polarization scheme~\cite{magic-1,magic-2,magic-3} 
for real emission of massless
gauge particles in such functions, it is natural to seek
further simplification and numerical stability in the virtual 
emission and re-absorption processes as well by exploiting the same scheme.
It is this that we pursue in what follows.\par
For the reader unfamiliar with the ``Chinese magic'' polarization scheme
for massless gauge bosons,
which is historically associated to the preprint in Ref.~\cite{magic-1},
the key observation is that the gauge invariance of the attendant massless 
gauge theory allows one to use an attendant 
set of polarization vectors which, when the
chiral forms of the respective spin $\frac{1}{2}$ charged particles'
wave functions are used, eliminate radiation from one entire side of
a charged line and, simultaneously, simplify considerably the calculation of the part of the amplitude 
that remains, almost like ``magic, hence the name.
This is possible because of a representation of the
respective polarization vector for helicity $\lambda_\gamma$
and 4-momentum $k_\gamma$,~$\epsilon_{\lambda_\gamma}^\mu$,
as a matrix element of the Dirac gamma matrix, $\gamma^\mu$, between 
the spinor of helicity $\lambda_\gamma$ and four-momentum $k_\gamma$,~$|k_\gamma \lambda_\gamma>$,
and the massless spinor state $<\rho \lambda_\gamma|, \; \rho^2=0$, up to a normalization factor,
so that the Chisholm identity (see eq.(\ref{chslm-1}) below) reduces the Feynman rule factor $\epsilon_{\lambda_\gamma}^{*\mu}\gamma_\mu$ at the 
respective interaction vertex
to the simple expression $2[|k_\gamma -\lambda_\gamma><\rho -\lambda_\gamma|+|\rho \lambda_\gamma><k_\gamma \lambda_\gamma|]$, up to the same normalization factor, which causes one side of a line of the real radiation terms to vanish if
$\rho$ is set equal to the external 4-momentum entering(leaving) that side of the respective line.
The remaining terms are then expressed in terms of simple spinor products which lend themselves
to easy evaluation~\cite{magic-1,magic-2,magic-3}. 
This gives a 'magically' shortened
expression compared the usual Cartesian representation of the
polarization vector with the squared amplitude modulus 
evaluated using traces over the
fermion lines. We illustrate this below here as well.\par
\begin{figure}[h]
\begin{center}
%x\epsfig{file=pent-1.eps,width=140mm}
\includegraphics[width=140mm]{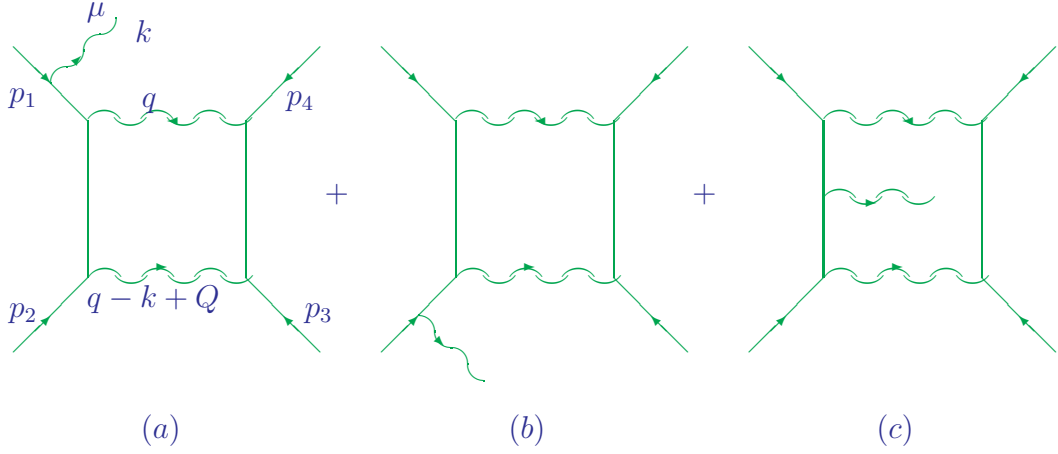}
\end{center}
\caption{\baselineskip=7mm  ISR 5-point function contributions
with fermion and vector boson masses $m_f,~m_B,\; f=1,2,\; B=V_1,V_2$ and with four momenta $p_i,~k$ as shown, with $Q\equiv p_1+p_2$. Radiation is shown from the initial state line with electric charge $Q_1e$ where $e$ is the electric charge of the positron -- here $p_1$ is the incoming fermion 4-momentum, $p_2$ is the incoming anti-fermion 4-momentum. When the quantum numbers allow it, the crossed graphs for the internal vector boson exchanges must be added to what we show
here.}
\label{fig1-5pt}
\end{figure}

Specifically, we will use the conventions of Ref.~\cite{ceex,gps}
for spinors and polarization vectors, which are derived from the
work of ~\cite{magic-1,magic-3}. The 5-pt function which we want to
analyze in these conventions as our prototypical example is shown in diagram (c) in Fig.~\ref{fig1-5pt}.
It has many applications
in collider precision physics. When combined with diagrams 1(a) and 1(b)
it generates a gauge invariant contribution to the ISR for $e^+e^-\rightarrow f\bar{f}+\gamma$,~$f\ne e$~\footnote{A numerical realization of the amplitude in Fig.~\ref{fig1-5pt} as it relates to bhabha scattering can be found in Ref.~\cite{actis}.}, for example, and it is a part of such
a contribution to $u\bar{u}\rightarrow \mu\bar{\mu}+G$ (in an appropriate color
basis), etc. 
%A numerical realization of the amplitude in Fig.~\ref{fig1-5pt}
%can be found in Ref.~\cite{actis}.
Such applications and their attendant phenomenology will be taken up elsewhere.~\cite{elswh}. Here, we focus on the use of Chinese magic in the loop integral in Fig. 1(c) to illustrate what simplifications are possible.\par
More precisely, by the standard methods, we need the following Feynman
integral representation of Fig. 1(c){\small
\begin{equation}
\begin{split}
{\cal M}^{(1c)}_{\lambda_1\lambda_2\lambda'_1\lambda'_2\lambda_\gamma}&=(2\pi)^4\delta(p_1+p_2-p'_1-p'_2-k){\cal C}\frac{\int d^4q}{(2\pi)^4}\frac{\bar{v}_{\lambda_2}\gamma^\beta(\not\!q+\not\!p_1-\not\!k+m_1)\not\!\epsilon^*_{\lambda_\gamma}(\not\!q+\not\!p_1+m_1)\gamma^\alpha u_{\lambda_1}}{((q+p_1-k)^2-m_1^2+i\epsilon)((q+p_1)^2-m_1^2+i\epsilon)}\\
&\quad \frac{\bar{u}'_{\lambda'_1}\gamma_\alpha(\not\!q+\not\!p'_1+m_2)\gamma_\beta v'_{\lambda'_2}}{((q+p_1+p_2-k)^2-M_{V_2}^2+i\epsilon)((q+p'_1)^2-m_2^2+i\epsilon)(q^2-M_{V_1}^2+i\epsilon)}+\ldots,
\end{split}
\label{eq-int1}
\end{equation}}
where we have defined massless limit coupling factor
\begin{equation}
{\cal C}={\cal C}(\{\lambda_i\},\{\lambda'_j\})=Q_1eG^2G'^2(v'_1+a'_1\lambda_2)(v_1-a_1\lambda_1)(v'_2+a'_2\lambda'_2)(v_2-a_2\lambda'_1)
\end{equation}
with the couplings $Q_1e,\; G\; \text{and}\; G'$ for the $\gamma,\; V_1\; \text{and}\; V_2$, respectively. In the usual Glashow-Salam-Weinberg-'t Hooft-Veltman~\cite{gsw} 
notation, v(a) represents
vector(axial-vector) coupling.
The ellipsis in (\ref{eq-int1}) represent the mass corrections 
needed to correct the massless limit used for ${\cal C}(\{\lambda_i\},\{\lambda'_j\})$. They are not necessary to illustrate our method and they will
be restored elsewhere~\cite{elswh}.
To get the loop integral in terms of Chinese magic, we take the following
kinematics as shown in Fig.~\ref{fig1-5pt}:
\begin{eqnarray}
p_1&=&(E,p\hat{z})\nonumber\\
p_2&=&(E,-p\hat{z})\nonumber\\
-p_4&=&(E',p'(\cos\theta'_1\hat{z}+\sin\theta'_1\hat{x}))\equiv p'_1\nonumber\\
k&=&(k^0,k(\cos\theta_\gamma\hat{z}+\sin\theta_\gamma(\cos\phi_\gamma\hat{x}+\sin\phi_\gamma\hat{y})))\nonumber\\
-p_4-p_3+k&=&p_1+p_2=(\sqrt{s},\vec{0})\nonumber\\
-p_3&\equiv& p'_2
\end{eqnarray}
with $k^0=k$, $\sqrt{s}=2E$. Here, we introduce the alternate notations
$p'_1=-p_4,\; p'_2=-p_3$ for cosmetic use entirely. We now introduce the 
two sets of magic polarization vectors associated to the two incoming lines:
\begin{equation}
%\begin{equation}
\label{gmapol}
  (\epsilon^\mu_\sigma(\beta))^*
     ={\bar{u}_\sigma(k) \gamma^\mu u_\sigma(\beta)
       \over \sqrt{2}\; \bar{u}_{-\sigma}(k) u_\sigma(\beta)},\quad
  (\epsilon^\mu_\sigma(\zeta))^*
     ={\bar{u}_\sigma(k) \gamma^\mu \umf_\sigma(\zeta)
       \over \sqrt{2}\; \bar{u}_{-\sigma}(k) \umf_\sigma(\zeta)},
\end{equation}
with $\beta^2=0$ and $\zeta$ defined in Ref.~\cite{ceex,gps}, 
so that all phase information is strictly known in our amplitudes: the two
choices for $\beta$ are such that its space-like components have the directions
of the two incoming beams in the initial state.
We take the  the basis of the 4-dimensional momentum space as follows:
\begin{equation}
\begin{split}
\ell_1&=(E,E\hat{z}),\;\ell_2=(E,-E\hat{z})\cr
\ell_3&=E\frac{<\ell_2+|\gamma^\mu|\ell_1+>}{\sqrt{2}<\ell_2-|\ell_1+>}\cr
      &=\quad \frac{-E}{\sqrt{2}}(\hat{x}+i\hat{y})\cr
\ell_4&=E\frac{<\ell_2-|\gamma^\mu|\ell_1->}{\sqrt{2}<\ell_2+|\ell_1->}\cr
      &=\quad \frac{E}{\sqrt{2}}(\hat{x}-i\hat{y})
\end{split}
\end{equation}
where we use the obvious equivalence $|\ell\sigma>=u(\ell)_\sigma$
in the notation of Refs.~\cite{magic-1,magic-2,magic-3,gps}.
The important point is that all four of these basis 4-vectors are light-like with $\ell_i^2=0, \; i=1,\cdots,4$.They 
therefore can participate in Chinese magic.
\par
To illustrate explicitly this latter point, consider the definite case 
$\lambda_1,\lambda_2,\lambda'_1,\lambda'_2,\lambda_\gamma =+,-,+,-,+$,
as all other choices for the helicities behave similarly.
We write the loop momentum as 
\begin{equation}
q=\alpha_i\ell_i
\end{equation}
with summation over repeated indices understood. The coefficients 
$\alpha_i$ are readily determined as
\begin{equation}
\begin{split}
\alpha_1&=\frac{q\ell_2}{2E^2}=({\frak D}_3-{\frak D}_2-s+2p_2k+M_{V_2}^2)/s\cr
\alpha_2&=\frac{q\ell_1}{2E^2}=({\frak D}_1-{\frak D}_0-M_{V_1}^2)/s\cr
\alpha_3&=\frac{q\ell_4}{E^2}=-\frac{q\ell^*_3}{E^2}=-\alpha^*_4\cr
\alpha_4&=-\frac{i}{\sqrt{2s}}[c_j{\frak D}_j+c_5M_{V_1}^2+c_6(M_{V_2}^2+2p_2k-s)+c_7(2kp_1)]
\end{split}
\end{equation} 
where we define the denominators as
\begin{equation}
\begin{split}
{\frak D}_0&=q^2-M_{V_1}^2+i\epsilon\cr
{\frak D}_1&=(q+p_1)^2-m_1^2+i\epsilon\cr
{\frak D}_2&=(q+p_1-k)^2-m_1^2+i\epsilon\cr
{\frak D}_3&=(q+p_1+p_2-k)^2-M_{V_2}^2+i\epsilon\cr
{\frak D}_4&=(q-p_4)^2-m_2^2+i\epsilon
\end{split}
\end{equation}
so that the expansion coefficients $\{c_j\}$ are
\begin{equation}
\begin{split}
c_0&=\csc\phi_\gamma(\frac{\csc\theta'_1 e^{i\phi_\gamma}}{\beta'_1E'_1}-\frac{\csc\theta'_1 e^{i\phi_\gamma}}{\beta'_1\sqrt{s}}+\frac{\csc\theta_\gamma}{\sqrt{s}}-\frac{\cot\theta'_1 e^{i\phi_\gamma}-\cot\theta_\gamma}{\beta_1\sqrt{s}})\cr
c_1&=\csc\phi_\gamma(\frac{\csc\theta'_1 e^{i\phi_\gamma}}{\beta'_1\sqrt{s}}-\frac{\csc\theta'_1}{\sqrt{s}}+\frac{\cot\theta'_1 e^{i\phi_\gamma}-\cot\theta_\gamma}{\beta_1\sqrt{s}}+\frac{\csc\theta_\gamma}{k^0})\cr
c_2&=\csc\phi_\gamma(\frac{-\csc\theta'_1 e^{i\phi_\gamma}}{\beta'_1\sqrt{s}}+\frac{\csc\theta_\gamma}{\sqrt{s}}+\frac{\cot\theta'_1 e^{i\phi_\gamma}-\cot\theta_\gamma}{\beta_1\sqrt{s}}-\frac{\csc\theta_\gamma}{k^0})\cr
c_3&=\csc\phi_\gamma(\frac{\csc\theta'_1 e^{i\phi_\gamma}}{\beta'_1\sqrt{s}}-\frac{\csc\theta_\gamma}{\sqrt{s}} -\frac{\cot\theta'_1e^{i\phi_\gamma}-\cot\theta_\gamma}{\beta_1\sqrt{s}})\cr
c_4&=-\csc\phi_\gamma\frac{\csc\theta'_1 e^{i\phi_\gamma}}{\beta'_1E'_1}\cr
c_5&=\csc\phi_\gamma(\frac{\csc\theta'_1 e^{i\phi_\gamma}}{\beta'_1E'_1}-\frac{\csc\theta'_1 e^{i\phi_\gamma}}{\beta'_1\sqrt{s}}+\frac{\csc\theta_\gamma}{\sqrt{s}}-\frac{\cot\theta'_1 e^{i\phi_\gamma}-\cot\theta_\gamma}{\beta_1\sqrt{s}})\cr
c_6&=\csc\phi_\gamma(\frac{\csc\theta'_1 e^{i\phi_\gamma}}{\beta'_1\sqrt{s}}-\frac{\csc\theta_\gamma}{\sqrt{s}}-\frac{\cot\theta'_1 e^{i\phi_\gamma}-\cot\theta_\gamma}{\beta_1\sqrt{s}})\cr
c_7&=-\csc\phi_\gamma\frac{\csc\theta_\gamma}{k^0}.
\end{split}
\end{equation}
Thus, the $\{c_j\}$ are determined explicitly by the cms kinematics that we 
use. The consequence to note is that the Chinese magic now carries over
to the loop variable via the identity 
\begin{equation}
\begin{split}
\not\!q&=\alpha_j\not\!\ell_j\cr
      &=\sum_{j=1}^2\alpha_j(|\ell_j+><\ell_j+|+|\ell_j-><\ell_j-|)\cr
      &+\alpha_3\frac{\sqrt{2}E}{<p_2-|p_1+>}(|\ell_2-><\ell_1-|+|\ell_1+><\ell_2+|)\cr
&+\alpha_4\frac{\sqrt{2}E}{<p_2+|p_1->}(|\ell_2+><\ell_1+|+|\ell_1-><\ell_2-|)\cr
      &\equiv \sum_{j=1}^2\alpha_j(|p_j+><p_j+|+|p_j-><p_j-|)\cr
      &+\alpha_3\frac{\sqrt{2}E}{<p_2-|p_1+>}(|p_2-><p_1-|+|p_1+><p_2+|)\cr
&+\alpha_4\frac{\sqrt{2}E}{<p_2+|p_1->}(|p_2+><p_1+|+|p_1-><p_2-|)\cr
       &\equiv \sum_{j=1}^2\alpha_j(|p_j+><p_j+|+|p_j-><p_j-|)\cr
      &+\tilde{\alpha}_3(|p_2-><p_1-|+|p_1+><p_2+|)\cr
&+\tilde{\alpha}_4(|p_2+><p_1+|+|p_1-><p_2-|)\cr
\end{split}
\label{eq-id1}
\end{equation}
where we work in the massless limit for this numerator algebra so that we take
$\ell_1\equiv p_1,\; \ell_2\equiv p_2$ in (\ref{eq-id1}). Here, we defined
as well
\begin{equation}
\begin{split}
\tilde{\alpha}_3&\equiv \alpha_3\frac{\sqrt{2}E}{<p_2-|p_1+>}=-\frac{\alpha_3}{\sqrt{2}}\cr
\tilde{\alpha}_4&\equiv \alpha_4\frac{\sqrt{2}E}{<p_2+|p_1->}= \frac{\alpha_4}{\sqrt{2}}\cr
\end{split}
\end{equation}
From the standpoint of efficient and numerically stable 
 MC event generator realization 
of the correction in Fig.~\ref{fig1-5pt}, the explicit form 
the $\alpha_j$ cannot be stressed too much.\par
Upon introducing the representation (\ref{eq-id1}) into the numerator, $N$, of
the integrand in (\ref{eq-int1}) we get, from the standard identities
\begin{equation}
\begin{split}
&\not\!\epsilon^*_{\lambda_\gamma}=\frac{\sqrt{2}}{<k -\lambda_\gamma|\ell_1 \lambda_\gamma>}[|\ell_1 \lambda_\gamma><k \lambda_\gamma|+|k -\lambda_\gamma><\ell_1 -\lambda_\gamma|],\cr
&\gamma^\rho<\ell_1\lambda|\gamma_\rho|\ell_2\lambda>=2[|\ell_1 -\lambda><\ell_2 -\lambda|+|\ell_2 \lambda><\ell_1 \lambda|],\cr
&\not\!\ell_1=|\ell_1 +><\ell_1 +|+|\ell_1 -><\ell_1 -|, 
\end{split}
\label{chslm-1}
\end{equation} 
the reduction
\begin{equation}
\begin{split}
N&= \frac{4\sqrt{2}}{<k -|p_1 +>}\big\{(A_1<p_2+|p'_1-><p'_2-|p_2+>+A_2<p_2+|p'_1-><p'_2-|p_1+>)\cr
&\qquad\qquad (A_3<p_2+|p'_1-><p'_1-|p_1+>+A_4<p_1+|p'_1-><p'_1-|p_1+>)\cr
&\qquad\qquad+ \tilde\alpha_4(A_1<p_2+|p_1-><p'_2-|p_2+>+A_2<p_2+|p_1-><p'_2-|p_1+>)\cr
&\qquad\qquad(A_3<p_2+|p'_1-><p_2-|p_1+>+A_4<p_1+|p'_1-><p_2-|p_1+>)\big\},
\end{split}
\label{num1}
\end{equation}
where we defined
\begin{eqnarray}
A_1&=&\tilde\alpha_4<p_1+|k->+\alpha_2<p_2+|k->,\nonumber\\
A_2&=&(1+\alpha_1)<p_1+|k->+\tilde\alpha_3<p_2+|k->\nonumber\\
A_3&=&\alpha_2<p_1-|p_2+>,\;\; A_4=\tilde\alpha_4<p_1-|p_2+>
\end{eqnarray}
for the magic choice $\beta=p_1$. Note that the 'magic' has killed all but one set
of the terms with three factors of the virtual momentum expansion
coefficients and that, in the numerator of the propagator (before)after the real emission vertex, it has eliminated the terms associated with 
($\not\!p_1$)$\not\!k$ as well as half of the terms in the
respective virtual momentum expansion in former case.
While we have eliminated a large fraction of the 
possible terms on the RHS of (\ref{num1}), one can ask how it
compares in length with what one would get from 
the usual approaches of taking traces on the fermion lines. To be specific, 
in the traditional method that leads to traces on fermion lines, 
one needs to compare the length of $2\Re {\cal M}^*_B{\cal M}^{(1c)}$
where ${\cal M}_B$ is the respective Born amplitude that would interfere
with the one-loop amplitude to create the one-loop correction to the
respective cross section. In the Chinese magic representation, we get immediately that only radiation from the anti-particle ($p_2$) incoming line contributes
with the simple result (repeated indices are summed and $s'=(p_1+p_2-k)^2$
as usual){\small
\begin{equation}
\begin{split}
{{\cal M}_B}_{+-+-+} &= (2\pi)^4\delta(p_1+p_2-p'_1-p'_2-k)\frac{2\sqrt{2}ieQ_1G_j^2(v'_j-a'_j)(v_j-a_j)<p'_2-|p_1+>}{<k-|p_1+><k-|p_2+>(s'-M_{V_j}^2+i\epsilon)}\\
&\;\;\;{\big[}<p_1-|p_2+><p_2+|p'_1->-<p_1-|k+><k+|p'_1->{\big]}
\end{split}
\label{eq-brn}
\end{equation}}
so that computing $2\Re {\cal M}^*_B{\cal M}^{(1c)}$ just involves multiplying $N$ in (\ref{num1}) by the complex conjugate of this simple expression and taking twice the real part. If we proceed with the usual trace on the fermion lines method,
one needs the trace of two sets of terms with 10 Dirac gamma matrices multiplied by a factor with the trace of 6 Dirac 
gamma matrices: this means one has $2\cdot9\cdot7\cdot 5\cdot 4$$\times$$5\cdot4 = 2520\times 20=50,400$ terms, each of which
requires Passarino-Veltman reduction of 3, 2, and 1 5-pt tensor integrals.
In Ref.~\cite{glver}, another approach that leads as well to traces over fermions is used in which one first expands the amplitude under study in a gauge invariant tensor basis with scalar coefficients and uses Chinese magic-type~\cite{magic-1,magic-2,magic-3} representations of the helicity states to express the attendant helicity amplitudes in terms of these invariant scalar coefficient functions. The key step is the use of projection operators, ${\cal P}(X)$ in the notation of Ref.~\cite{glver}, which project out the scalar coefficient $X$. To compare with our approach, we observe the following: the Born amplitude tensor structure is one of the tensor structures in the respective expansion basis and to project its coefficient the respective projection operator evaluates a linear combination of the trace on the fermion lines of the hermitian conjugate of this Born level tensor structure in product with the Feynman amplitude and the traces on the fermion lines of the hermitian conjugates of the other tensor structures in product with the same amplitude. Thus, our counting of terms given for the evaluation of $2\Re {\cal M}_B^*{\cal M}^{(1c)}$ using the traditional traces on fermion lines gives a lower limit to the number of terms that would be generated by the methods of Ref.~\cite{glver} for our calculation\footnote{For example, let us take the example discussed in Ref.~\cite{glver}, using their notation, of $q(p_2,\lambda_2)\;\bar{q}(p_1,\lambda_1)\rightarrow \gamma(p_3,\lambda_3)\;\gamma(p_4,\lambda_4)$, where we focus just on the one-loop correction from the Gross-Wilczek-Politzer~\cite{qcd} QCD theory with direct analysis for the respective 4-pt box graph in which a gluon is exchanged between the incoming quark$(q)$ anti-quark$(\bar{q})$ pair ``before'' they annihilate to the two photons. For the helicities $+-++$ for the quark, anti-quark, $\gamma(p_3)$, $\gamma(p_4)$, respectively, the helicity amplitude is proportional to the $A_{11}$ scalar coefficient in Ref.~\cite{glver}. Evaluation of the projection operator for $A_{11}$ on the box graph requires the trace for a product of 12 Dirac gamma matrices, which generates $11\cdot 9\cdot 7\cdot 5\cdot 3=10,395$ terms, and this has to be done $5$ times (there are five scalar coefficients) for a total of $51,975$ terms. This is just a 4-pt function. The same calculation using our methods generates a formula smaller in length 
than that in eq.(\ref{eq-4-pt}) in the text.}.
Looked at this way, we can appreciate better the great simplification that (\ref{num1}) represents.
It follows that this form of $N$ in (\ref{num1}) has efficiently reduced
the problem of reduction of the 5-pt function with three, two and one tensor indices(index) 
in the Passarino-Veltman formalism to the problem of 
a single scalar 5-pt function and lower 4, 3 and 2 point functions with
the coefficients already explicitly expressed in terms of the cms kinematic variables that are so crucial to efficient MC event generation. Efficient MC
event generator realization of the latter functions is known~\cite{bkj,bkh,alibaba,sjward,ceex,bdp,bdh,bvnb,bmr,bardin,hollik,bwl,fleis-jeg,bohm,melles,mcnlo,powheg}, where it is understood that 
one uses the results
in Ref.~\cite{melrose,nto4-1,nto4-2,nto4-3} to express the scalar 5-pt function in terms of scalar
4-pt functions using our explicit kinematics above. 
These last remarks are made more manifest when one notes the introduction of the result
for $N$ in (\ref{num1}) into the integral in (\ref{eq-int1}) leads to the integrals 
\begin{equation}
\frac{\int d^4q}{(2\pi)^4}\frac{{\frak D}_i{\frak D}_j{\frak D}_k;{\frak D}_i{\frak D}_j;{\frak D}_j;1}{{\frak D}_0{\frak D}_1{\frak D}_2{\frak D}_3{\frak D}_4},\;\; i,j,k=0,\cdots,4
\label{eq-int2}
\end{equation}
all of which are known from the lower point functions we advertised when the results
for the representation of the scalar 5-point function in terms of 4-point functions
in Refs.~\cite{nto4-1,nto4-2,nto4-3} are used\footnote{Since the integral in (\ref{eq-int1})
is manifestly UV finite, we do not need to specify what regularization is used for the two point functions because only UV finite combinations of them can occur here while the wave functions are all in 4-dimensional Minkowski space. Note also that the standard trace over fermion lines would also lead to 
results equivalent to that in (\ref{eq-int2}) but as we have seen above
it would necessitate evaluation and simplification
of much longer expressions in general to compute the attendant
transition rate for the process.}. We get a bonus:  
{\it no evaluation of wave functions at complex momenta} is required here. 
What we have done is rigorously
a result of Lagrangian quantum field theory and it therefore can serve as a cross check on methods that may not obviously so be.
Evidently, the method we have illustrated can be used for any higher point function. \par
At this point, while we have shortened considerably the respective amplitude
and have removed the Gram determinant type factors in the tensor reductions,
we are still subject to the Gram determinant-type
denominator factors
in the results in Refs.~\cite{nto4-1,nto4-2,nto4-3} for the representation
of the 5-point scalar function in terms of 4-point scalar functions.
We have found that these are in general still too numerically unstable for
realization in the amplitude-based exact resummation MC event generators
such as those in Refs.~\cite{sjward}. Thus, we replace the representation
from Refs.~\cite{nto4-1,nto4-2,nto4-3}
of the needed 5-point scalar function here as follows.
\par
We start from the basic identity 
\begin{equation}
\begin{split}
q^2&={\frak D}_0 +M_{V_1}^2-i\epsilon\\
   &=(\alpha_i\ell_i)^2\\
   &=2\alpha_1\alpha_2\ell_1\ell_2+2\alpha_3\alpha_4\ell_3\ell_4\\
   &=s\alpha_1\alpha_2 +\frac{s}{2}\alpha_3\alpha_4.
\end{split}
\end{equation}
Dividing by ${\frak D}_0\cdots {\frak D}_4$ and integrating over $d^4q$
we arrive at the following representation of the required scalar 5-point function (we use the notation of Ref.~\cite{nto4-3} for $E_0$ itself):{\small 
\begin{equation}
\begin{split}
E_0(\bar{p}_1,\bar{p}_2,\bar{p}_3,\bar{p}_4,\bar{m}_0,\bar{m}_1,\bar{m}_2,\bar{m}_3,\bar{m}_4)
%|_{\bar{p}_1=p_1,\bar{p}_2=p_1-k,\bar{p}_3=p_1+p_2-k,\bar{p}_4=p'_1,\bar{m}_0=M_{V_1},\bar{m}_1=m_1,\bar{m}_2=m_1,\bar{m}_3=M_{V_2},\bar{m}_4=m_2}
&={\Big\{}-D_0(0)+\frac{1+\beta^2}{2s\beta^2}[C_0(13)-C_0(12)-C_0(03)+C_0(02)\\
&\;\;+(M_{V_2}^2-s+2p_2k)(D_0(1)-D_0(0))-M_{V_1}^2(D_0(3)-D_0(2))]\\
&\;\;-\frac{1-\beta^2}{4s\beta^2}[\Delta r_{1,0}(D_0(1)-D_0(0))+2\Delta\bar{p}_{1,0}(D_{11}(1)\bar{\bar p}(1)_1\\
&\;\;-D_{11}(0)\bar{\bar p}(0)_1+D_{12}(1)\bar{\bar p}(1)_2-D_{12}(0)\bar{\bar p}(0)_2+D_{13}(1)\bar{\bar p}(1)_3\\
&\;\;-D_{13}(0)\bar{\bar p}(0)_3)-D_0(1)\bar{\bar p}(1)_4+D_0(0)\bar{\bar p}(0)_4\\
&\;\; -2M_{V_1}^2(D_0(1)-D_0(0))+\Delta r_{3,2}(D_0(3)-D_0(2))\\ 
&\;\;+2\Delta\bar{p}_{3,2}(D_{11}(3)\bar{\bar p}(3)_1-D_{11}(2)\bar{\bar p}(2)_1+D_{12}(3)\bar{\bar p}(3)_2-D_{12}(2)\bar{\bar p}(2)_2\\
&\;\;+D_{13}(3)\bar{\bar p}(3)_3-D_{13}(2)\bar{\bar p}(2)_3 -D_0(3)\bar{\bar p}(3)_4+D_0(2)\bar{\bar p}(2)_4)\\
&\;\;+2(M_{V_2}^2-s+2p_2k)(D_0(3)-D_0(2))]-\frac{1}{4}{\Big[}\sum_{j=0}^{4}|c_j|^2(C_0(j,j+1)\\
&\;\;+\Delta r_{j,j+1}D_0(j)+2\Delta\bar{p}_{j,j+1}(D_{11}(j)\bar{\bar p}(j)_1+D_{12}(j)\bar{\bar p}(j)_2\\
&\;\;+D_{13}(j)\bar{\bar p}(j)_3-D_0(j)\bar{\bar p}(j)_4))+2(\sum_{i<j}^{4}\Re(c_ic_j^*)C_0(ij)\\
&\;\;+\sum_{j=0}^4\Re(c_j(c_5^*M_{V_1}^2+c_6^*(M_{V_2}^2-s+2p_2k)+c_7^*(2kp_1)))D_0(j)){\Big]} {\Big\}}/C_{E_0},
\end{split}
\end{equation}}
where we have the identifications{\small
$$\bar{p}_1=p_1,\;\bar{p}_2=p_1-k,\;\bar{p}_3=p_1+p_2-k,\;\bar{p}_4=p'_1,\;\bar{m}_0=M_{V_1},\;\bar{m}_1=m_1,\;\bar{m}_2=m_1,\;\bar{m}_3=M_{V_2},\;\bar{m}_4=m_2$$}
and where the coefficient $C_{E_0}$ is given by{\small
\begin{equation}
\begin{split}
C_{E_0}&=M_{V_1}^2-i\epsilon+\frac{1+\beta^2}{2\beta^2s}M_{V_1}^2(M_{V_2}^2-s+2p_2k)+
\frac{1-\beta^2}{4\beta^2s}(M_{V_1}^4+(M_{V_2}^2-s+2p_2k)^2)\\
&\;\;+\frac{1}{2}\Re[c_5c_6^*M_{V_1}^2(M_{V_2}^2-s+2p_2k)+c_5c_7^*M_{V_1}^2(2kp_1)+c_6c_7^*(M_{V_2}^2-s+2p_2k)(2kp_1)]+\frac{1}{4}[|c_5|^2M_{V_1}^4\\
&\;\;+|c_6|^2(M_{V_2}^2-s+2p_2k)^2+|c_7|^2(2kp_1)^2].
\end{split}
\end{equation}}
We have here used a combination of the notation from Ref.~\cite{pass-velt,nto4-2,nto4-3} so that the definitions which follow should hold true:{\small
\begin{equation}
\begin{split}
{\frak D}_j&=(q+\bar{p}_j)^2-\bar{m}_j^2+i\epsilon = q^2+2q\bar{p}_j+\bar{p}_j^2-\bar{m}_j^2+i\epsilon\equiv q^2+2q\bar{p}_j+r_j,\nonumber\\
\Delta r_{i,j}&\equiv r_i-r_j,\nonumber\\
\Delta\bar{p}_{i,j}&\equiv \bar{p}_i-\bar{p}_j,\nonumber\\
D_0(j) &\equiv \text{4-point scalar function obtained from 5-point scalar function by omitting denominator}\; {\frak D}_j,\nonumber\\
C_0(i,j)&\equiv \text{3-point scalar function obtained from 5-point scalar function by omitting denominators}\; {\frak D}_i\\
&\;\;\quad  \text{and}\; {\frak D}_j,\; i\ne j,
\end{split}
\end{equation}}
where  we also follow the Passarino-Veltman\protect\cite{pass-velt} notation of the 4-point one-tensor integral, $D_\mu(j)$, obtained by omitting denominator ${\frak D}_j$ from the corresponding 5-point one-tensor integral with
$$D_\mu(j)\equiv D_{11}(j)\bar{\bar p}(j)_1+D_{12}(j)\bar{\bar p}(j)_2+D_{13}(j)\bar{\bar p}(j)_3-D_0(j)\bar{\bar p}(j)_4$$
, where the 4-vectors $\{\bar{\bar p}(j)_i\}$ are then 
determined in accordance with Ref.~\cite{pass-velt}, with the understanding
that $\bar{\bar p}(j)_4$ is only non-zero if it is necessary to shift
the q integration variable by it to reach the standard form of the 
respective Passarino-Veltman representation.
This expression for $E_0$ does not have problems with Gram determinant type
denominators.\par
To further exhibit the magic in the polarization vector spinor representation
under display here, we record as well the results for Fig.~(1a) and (1b)
that one needs to add to our result for Fig.~(1c) to get a gauge invariant result:{\small
\begin{equation}
\begin{split}
{\cal M}^{(1a)}_{+-+-+}&= 0, \;\;\text{by 'magic'}\\
{\cal M}^{(1b)}_{+-+-+}&=(2\pi)^4\delta(p_1+p_2-p'_1-p'_2-k)\frac{4\sqrt{2}{\cal C}}{<k-|p_1+><k-|p_2+>}\frac{\int d^4q}{(2\pi)^4}\frac{N'}{{\frak D}_O{\frak D}_1{\frak D}_3{\frak D}_4}
\end{split}
\label{eq-4-pt}
\end{equation}}
where the numerator $N'$ is given by{\small
\begin{equation}
\begin{split}
N'&=\left(<p'_2-|p_1+>{\frak a}_1+<p'_2-|p_2+>{\frak b}_1\right)(<p_1-|p_2+><p_2+|p'_1->\\
&\;\;-<p_1-|k+><k+|p'_1->)+\left(<p'_2-|p_1+>\bar{\frak a}_1+<p'_2-|p_2+>\bar{\frak b}_1\right)\\
%&\;\;[(-2p_1p_2)\tilde\alpha_4+<p_1-|k+>(\alpha_2<k+|p_2->+\tilde\alpha_4<k+|p_1->)]
&\;\;[(-2p_1(p_2-k))\tilde\alpha_4+\alpha_2<p_1-|k+><k+|p_2->]
\end{split}
\end{equation}}
with the definitions{\small
\begin{equation}
\begin{split}
{\frak a}_1&=(1+\alpha_1)(2p_1p'_1)+\alpha_3<p_2+|p'_1-><p'_1-|p_1+>\\
{\frak b}_1&=\alpha_2<p_2+|p'_1-><p'_1-|p_1+>+\tilde\alpha_4(2p_1p'_1)\\
\bar{\frak a}_1&=<p_1-|p_2+>[(1+\alpha_1)<p_1+|p'_1->+\tilde\alpha_3<p_2+|p'_1->]\\
\bar{\frak b}_1&=<p_1-|p_2+>[\alpha_2<p_2+|p'_1->+\tilde\alpha_4<p_1+|p'_1->].
\end{split}
\end{equation}}
Again, this gives immediate reduction to the known scalar functions
with considerable reduction in the number of terms requiring evaluation
compared to the usual trace over fermion lines method when one computes the
respective contribution to $2\Re {\cal M}^*_B{\cal M}^{1b}$.
The complete phenomenology of our results for the process in Fig.~\ref{fig1-5pt} will appear elsewhere~\cite{elswh}.\par
It is important to explain the difference between what we have done here and 
what was done in Refs.~\cite{z-dix-kos,opp,pitt2,vanH1}. We do this in 
turn in a somewhat reverse chronological order. 
In Ref.~\cite{pitt2}, the representation of the 
loop variable in a basis of light-like 4-vectors is used to construct a 
recursion relation between one-loop n-point tensor integrals of differing 
rank whereas in Ref.~\cite{vanH1} the spinor representation of the external 
tensor coefficient of a massless n-point tensor one-loop integral is used to 
reduce the rank of that integral iteratively to allow numerical implementation,
using Dirac matrix methods. In both cases, the square roots of the Gram 
determinants appear in the denominators of the resulting representations. 
In our approach, explicit kinematics allows direct Chinese magic action in the 
complete amplitude contribution's evaluation directly to the lower 
point functions without Gram determinant factors to be computed 
in our denominators. No iteration is necessary and Chinese magic reduces 
considerably the number of terms in our final result. Such action is not 
present in Refs.~\cite{pitt2,vanH1}. In Refs.~\cite{opp}, the 
representation of the n-point amplitude at one-loop starts 
from its integrand $N(q)/({\frak D}_0\cdots {\frak D}_{n-1})$ 
with an expansion of the numerator $N(q)$ in powers of the denominators 
$\{{\frak D}_j\}$ with coefficients that split into a part that is independent of 
$q$ and a part that integrates to zero with the understanding that the 
integration measure is in general in $d$ dimensions whereas the function $N(q)$
is defined for $q$ in 4-dimensions. We refer to this representation as the OPP 
representation after the authors in the first paper in Refs.~\cite{opp}. 
Various methods for adding in the so-called missing rational terms generated 
by the mismatch between the 4-dimensional q in N and the d-dimensional 
$\bar{q}$ in the $\{{\frak D}_j\}$ are given in Refs.~\cite{opp}, including the 
generalized d-dimensional unitarity that treats the full d-dimensional 
unitarity realization of the OPP representation. In all of these works, 
$N(q)$ or $N(\bar{q})$ is treated as a given and no procedure for exploiting 
Chinese magic to simply it at the loop momentum level is considered. Moreover, 
the need to add in rational terms is an essential part of the procedure, 
whereas, as we see in our result (\ref{num1}), we do not have such an issue 
in our approach -- we get the complete answer with methods that operate
entirely in 4-dimensions\footnote{If one wants to apply our method to lower point amplitudes that are UV divergent, in renormalizable theories one should use the known counter-terms for those divergences to render the amplitudes finite first and then apply our 4-dimensional methods to the UV finite subtracted amplitudes.}. More importantly, 
inverse powers of Gram-type determinants 
appear in the coefficients in the representation of $N$
so that issues of numerical stability obtain, whereas as we show above our 
approach does not lead to such factors so that it should be more stable. 
Finally, the procedure for determining the coefficients in the representation 
of $N$ involves solving the algebraic problem for the q values at which 4 , 
then 3, then 2, and finally 1 of the $\{{\frak D}_j\}$ vanish(es). This means that, in 
general, complex values of q are required and this forces the evaluation 
of $N(q)$ at such unphysical 4-momenta. Our approach avoids this issue 
altogether as we carry our entire calculation out in the 4-dimensional real 
virtual loop momentum space. We then provide a completely physical cross 
check on the methods in Refs.~\cite{opp}. Similarly, the approach in 
Refs.~\cite{z-dix-kos} also takes the integrand as a given and constructs 
the respective amplitude from unitarity-based on-shell (recursion) relations, 
where the authors in Refs.~\cite{z-dix-kos} are able to get both the 
cut-constructable and the rational parts of the amplitudes with such methods. 
Again, there is no exploitation of Chinese magic to simply the amplitude at 
the loop variable level, the amplitude construction uses 4-particle cuts that 
have in general complex 4-momenta as their solutions so that 
wave functions are 
evaluated at such unphysical momenta, and the solution of these on-shell 
relations generally introduces troublesome kinematic factors in the 
denominators of the representation so that numerical stability cannot be 
assured. Our approach avoids all of these problems and affords again a 
completely physical cross check on this approach as well. 
\par  
The complete analytical result for the amplitude in Fig.~\ref{fig1-5pt} will 
be presented elsewhere~\cite{elswh}. Here, we have shown that the use of 
Chinese magic in the virtual loop momentum can reduce considerably the amount 
of algebra required for stable, efficient, manifestly physical computation of higher point virtual 
corrections with general mass scales, as they are needed for exact 
amplitude-based resummed MC event generator realization.\par

\section*{Acknowledgments}

We thank Prof. S. Yost and Dr. S. Majhi for useful discussions. We also thank 
Prof. Ignatios Antoniadis for the support and kind hospitality of the CERN 
TH Unit while this work was completed.

\end{document}